\documentclass[12pt]{iopart}

\usepackage{graphicx}

\begin{document}

\title{Metallic properties of magnesium point contacts}

\author{R H M Smit$^1$, A I Mares$^1$\footnote{Present address: Nederlands
Meetinstituut, PO-Box 654, NL-2600~AR, Delft, The Netherlands}, M
H\"afner$^{2,3}$, P Pou$^3$, J C Cuevas$^3$ and J M van Ruitenbeek$^1$}

\address{$^1$ Kamerlingh Onnes Laboratorium, Leiden University, PO~Box 9504,
\mbox{NL-2300~RA Leiden}, The Netherlands}
\address{$^2$ Institut f\"ur Theoretische Festk\"orperphysik and DFG-Center for    
Functional Nanostructures, Universit\"at Karlsruhe, D-76128 Karlsruhe, Germany}
\address{$^3$ Departamento de F\'{\i}sica Te\'orica de la Materia Condensada,
Universidad Aut\'onoma de Madrid, E-28049 Madrid, Spain}

\ead{ruitenbeek@physics.leidenuniv.nl}

\begin{abstract}
We present an experimental and theoretical study of the conductance and
stability of Mg atomic-sized contacts. Using Mechanically Controllable
Break Junctions (MCBJ), we have observed that the room temperature conductance
histograms exhibit a series of peaks, which suggests the existence of a shell
effect. Its periodicity, however, cannot be simply explained in terms of
either an atomic or electronic shell effect. We have also found that at room
temperature, contacts of the diameter of a single atom are absent. A possible
interpretation could be the occurrence of a metal-to-insulator transition as the
contact radius is reduced, in analogy with what it is known in the context of Mg
clusters. However, our first principle calculations show that while an infinite
linear chain can be insulating, Mg wires with larger atomic coordinations, as in
realistic atomic contacts, are always metallic. Finally, at liquid helium
temperature our measurements show that the conductance histogram is dominated
by a pronounced peak at the quantum of conductance. This is in good agreement
with our calculations based on a tight-binding model that indicate that the
conductance of a Mg one-atom contact is dominated by a single fully open
conduction channel.
\end{abstract}

\submitto{\NJP}

\maketitle

\section{Introduction}\label{sec:intro}

The understanding of conductance at the very small scale of a single atom
has advanced greatly over the last decade due to a joined effort from both
experiment and theory~\cite{agrait2003}. In this regime the total conductance,
$G$, is described by the Landauer formula: $G = \mbox{G}_0 \sum_{i=1}^{N}
\tau_i $, where G$_0 = 2 e^2 /h$ is the quantum of conductance and $\tau_i$
is the value of transmission for the i-th electronic mode (channel). For many
metals the conductance of a one-atom contact does not match the value of a
single quantum of conductance. Values in excess of $2e^2/h$, found for
transition metal contacts, indicate that the conductance has to be determined
by multiple channels. A pivotal contribution in the understanding of these
single-atom contacts was given by Scheer \textit{et al.}~\cite{scheer1998}.
They showed that the maximum number of modes to be considered was limited
by the number of valence orbitals of the central atom~\cite{cuevas1998a}.

In the case of \textit{s}$^{1}$ metals, such as the alkali and noble metals,
the conduction is given by one single electronic mode. Experimentally, the
transmission of this ``channel'' is found to be close to one, resulting in
a conductance close to $2 e^2 /h$ or a resistance of 13~k$\Omega$. For
\textit{sp} metals the total conduction is governed by three channels, but
this does not lead to a conductance of 3~G$_0$, since the electronic states
in general have a transmission $\tau < 1$. For the transition metals the
number of conduction channels even increases to five, due to the \textit{d}
orbitals.

When we shift our attention to Mg, however, it is not so easy to predict
what the conductance through the single atom will be. An isolated atom of
this \textit{s}$^{2}$ metal has the \textit{s}-shell fully filled, but the
\textit{p}-shell still empty. It is therefore not straightforward to predict
which electronic states are of importance to the conductance of a single-atom
contact. A comparable electronic structure can be found in Zn, where the
4\textit{s}-shell is completely filled, while the 4\textit{p}-shell is
empty. H\"afner~\textit{et al.}~\cite{hafner2004} have shown that the
conductance of a single-atom contact in this case was between 0.8 and 1.0~G$_0$,
where the transmission was largely given by a single channel.

The fact that bulk Mg is a metal, results from the hybridization of the
3\textit{s} and the 3$\textit{p}$ bands that originates from the large
coordination number (12 in a hcp structure). However, the situation might be
different when reducing this coordination. Indeed, it is well known that Mg
clusters exhibits unusual properties (see reference~\cite{issendorff2005} and
references therein). In particular, it has been shown that the
nonmetallic-to-metallic transition in Mg clusters is non-monotonic and clearly
slower than, for instance, in alkali metals~\cite{thomas2002}. This is due to
the dramatic change in the electronic structure of Mg with the number of
interacting atoms. In the extreme case of a dimer, the bonding is due to weak
van der Waals interactions. In the thinning of an atomic contact there is a
progressive reduction of the coordination of the atoms and, with it, the
hybridization. In this sense, the first question we want to address in this work
is whether this reduction is enough to induce a metal-to-insulator transition
for small Mg contacts.

Besides the cross-over from insulating to metallic behaviour, cluster
experiments also demonstrated a shell structure~\cite{thomas2002}. One
speaks of a shell structure if the mass spectrum shows a periodicity as a
function of the radius~\cite{martin1996}. The first observation of a shell
structure in magnesium clusters was found to be due to the closing of facets
and the crystalline arrangement~\cite{martin1991}. The hexagonally close-packed
lattice leads to the periodic occurrence of highly stable icosahedrons. Later
experiments performed with magnesium in supercold helium droplets showed shell
structure due to the delocalized electrons that display electronic level
bunching~\cite{diederich2001}.

Analogies to cluster shell structures were already found in quantum point
contacts of alkali metals~\cite{yanson2001} and subsequent work showed that
these are not limited to this chemical group~\cite{mares2005,mares2007}. For
metallic contacts to demonstrate shell structures the atoms require sufficient
thermal energy in order to find the local minima in energy. For sodium the
necessary temperature for observing the shell effects was found to be around
80~K~\cite{yanson1999}. Since magnesium has a much higher melting temperature
(922~K versus 371~K for Na), one expects the necessary temperature in this
case to be above 200~K. The second question we want to address, therefore, is
whether magnesium point contacts demonstrate shell structures at room
temperature. The fact that small clusters are insulating, while small contacts
have their electronic structure influenced by the leads, makes the comparison
between the two manifestations of the shell structures all the more interesting.

The rest of the paper is organized as follows. In the next section we describe
the experimental technique used in this work to study the mechanical and
transport properties of Mg atomic-sized contacts. Then, in \sref{sec:troom} we
discuss our observations at room temperature, which suggest the existence of a
shell effect. \Sref{sec:theory} is devoted to the theoretical analysis of Mg
nanocontacts. In particular, we present first principle calculations of the
electronic structure of Mg infinite wires of different thicknesses as well as
conductance calculations for Mg one-atom contacts based on a tight-binding
model. In \sref{sec:tlow} we discuss our experimental results for the
conductance of Mg point contacts at liquid helium temperature. Finally, we
summarize in \sref{sec:conc} the main conclusions of our work.

\section{Experimental Technique}\label{sec:tech}

In order to investigate Mg contacts experimentally, we have used Mechanically
Controllable Break Junctions (MCBJ) at both room and liquid helium temperature.
We start with a magnesium wire (purity better than 99.9 \%) of 125~$\mu$m
diameter and about 15~mm length, which we give a small incision in the middle.

For the low temperature measurements, the wire is glued on top of an insulating
Kapton layer, that covers a phosphor bronze bendable substrate. The glue
(Stycast Epoxy 2850FT with curing agent 24LV) is positioned on either side
and as close to the incision as possible. The bendable substrate is clamped
in a three-point bending configuration inside a vacuum pot which is brought to a
pressure $< 10^{-5}$~mbar. The pumping is done by an oil-free diaphragm pump
in combination with a turbo molecular pump to reduce possible contamination,
especially by hydrogen. This is important considering the strong chemical
affinity of magnesium to hydrogen~\cite{sakintuna2007}. After pumping, the
vacuum pot is submerged in liquid helium, cooling the sample down to a
temperature close to 4.2~K. An important benefit of this technique is
that it maintains the sample in a cryogenic vacuum and minimizes the
possibility of contamination. Using a mechanical axis for the coarse movement,
the wire is broken at the incision. In this way the two electrodes are formed
by freshly exposed surfaces, which is critical for the study of reactive
materials such as magnesium. A stacked piezo element is used for the fine
adjustments of the contact diameter during the experiment. By relaxing the
bending of the substrate the two clean surfaces can be brought into contact
again, adjusting the size of the contact to the level of single atoms.

The set-up we use to study magnesium at room temperature is comparable to the
one used at low temperatures~\cite{mares2004}. Instead of one substrate, we
now use two, fixed in line with a small gap as separation. In this way, both
the substrates can act as separate electrodes, avoiding the need for insulation
of the wire. The magnesium wire is stretched across the gap at the top and
fixed by small clamps on either side. One thus avoids the use of glue, which
can cause problems in the combination with ultrahigh vacuum (UHV). The piezo
is positioned below the gap and presses the two substrates upward and outward,
effectively breaking the wire. The ratio between the upward movement of the
piezo and the outward stretching of the sample wire (commonly called the
reduction ratio) is comparable to the low temperature configuration, resulting
in similar stability. The sample holder with the wire is mounted on a stainless
steel insert, which is inserted from the top into the UHV chamber. The chamber
is pumped down to $10^{-7}$~mbar with a turbo pump, after which it is baked at
450~K. In combination with an ion pump the pressure can be brought down to $1
\times 10^{-10}$~mbar. During the measurement, however, the turbo pump is shut
down to minimize the mechanical vibrations coupling to the sample. The base
pressure during the experiments is $4 \times 10^{-10}$~mbar.

The level of contamination in both these setups is too low to have an influence
on the results. This can be concluded from the data being reproducible during a
period of several days. Since the sample configuration is comparable in both
setups their difference is limited to the temperature. The surface diffusion of
atoms at low temperatures is frozen out, forcing the contact to the nearest low
energy state in configuration space. At room temperature, however, thermal
diffusion of the atoms allows the contact to probe a wider range of
configurations. Subtle energy differences such as the one caused by the shell
effect are then probed more effectively.

\section{Room temperature results}\label{sec:troom}

\begin{figure}
\centerline{\includegraphics[width=5in]{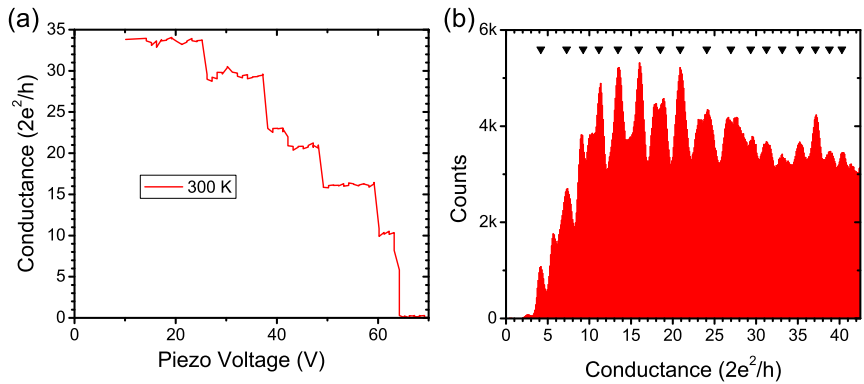}}
\caption{(a) The conductance evolution during the stretching of a magnesium
contact at room temperature in UHV. The conductance was measured at a bias
voltage of 20~mV. (b) A conductance histogram resulting from $1.2 \times
10^4$ of these traces. The resolution of this histogram was set to 10 bins per
G$_0$. The conductance peaks, obtained by an automatic procedure, are indicated
by arrows at the top of the panel.}
\label{fig:hit}
\end{figure}

The evolution of conductance when breaking the wire at room temperature has
a characteristic that can be seen in the example of \fref{fig:hit}(a). We
see a step-like lowering of the conductance towards zero with stretching of the
wire. Relatively stable intervals or plateaus occur when the mechanical tension
on the contact gradually builds up. When this tension is released by a
mechanical reconfiguration of the contact, its diameter and conductance are
reduced stepwise~\cite{rubio1996}. The plateaus in conductance appear flat and
the jumps during the mechanical reconfigurations are large. This is an
indication that some values of conductance are stabilized when they are related
to structures with a local minimum in the energy.

In order to verify that these plateaus occur at reproducible values we use
a statistical description by means of a conductance histogram~\cite{krans1995}.
This histogram is constructed by dividing the conductance interval of interest
in equal sub-intervals or bins. For each measured data point we determine the
corresponding bin, resulting in a probability distribution for the conductance
during breaking. An example of such a distribution or histogram for Mg is
given in \fref{fig:hit}(b). We varied the speed of the electrode separation
between  $1.0$ and $5.0 \times 10^1$~nm/s, but this had no significant effect
on the conductance histograms.

The result in \fref{fig:hit} indeed shows a series of peaks at higher
conductance values. The positions of these peaks were determined automatically
by a numerical procedure registering changes in the sign of its numerical
derivative. The obtained values are indicated by arrows in the same figure.
These could indicate the presence of a shell structure, but in order to
investigate whether the peaks in \fref{fig:hit} are indeed periodic in the
radius we need to plot the histogram as a function of radius, $R$. Since the
range of conductances in this graph corresponds to the ballistic conductance
regime we can obtain the $R$ from $G$ via the corrected Sharvin formula
\begin{equation}\label{eqn:sharvin}
 G \approx \mbox{G}_{0} \left[ \left( \frac{k_{F} R}{2} \right)^{2} -
 \frac{k_{F}R}{2}+\frac{1}{6}+ \dots \right]
\end{equation}
where $k_F$ represents the Fermi wave vector~\cite{torres1994,hoppler1998}.

\begin{figure} 
\centerline{\includegraphics[width=3in]{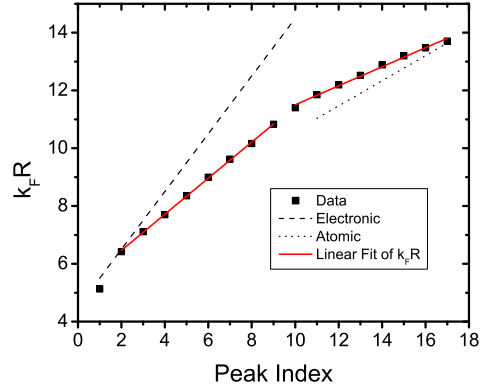}}
\caption{Graph showing the value for $k_FR$ obtained for the conductance peaks
in \fref{fig:hit}. The data points are successfully described by two linear
fits, indicating a periodicity of 1.6~$(k_F R)^{-1}$ at smaller contact
diameters and 3.2~$(k_F R)^{-1}$ at bigger contact diameters. Two additional
lines indicate the expected periodicity for atomic (dotted) and electronic
(dashed) shell effect as described in the text.}
\label{fig:kfrg}
\end{figure}

For the specific case of \fref{fig:hit} the calculated radius for the
peaks is plotted in \fref{fig:kfrg}. The obtained data points are clearly
described by two linear fits, indicating a crossover between two shell
effects with different periods~\cite{yanson2001}. For this specific histogram
the period changes from 1.6~$(k_F R)^{-1}$ for small contact diameters to
3.2~$(k_F R)^{-1}$ for larger contact diameters.

For comparing these values to the periodicity expected for shell structure due
to atomic packing, one starts from a simple packing model for a close-packed
hexagonal lattice and increases the contact diameter facet by facet. One then
obtains a periodicity of 2.3~$(k_F R)^{-1}$, which is only slightly influenced
when considering other crystal structures such as face (body) centred cubic. In
\fref{fig:kfrg} this period is represented by the dotted line, clearly
deviating from the obtained data.

Possible electronic shell structures have been studied for metallic point
contacts as well. Both, calculations using jellium models~\cite{ogando2002}
and nanoscale free-electron models~\cite{urban2006} applied to aluminium
point contacts~\cite{mares2007} give frequencies up to 1~$(k_F R)^{-1}$. As the
frequencies for Mg are suggested to be the same as for Al~\cite{urban2006}, we
represent this period by the dashed line in \fref{fig:kfrg}. Since none of the
lines provides a good description of the data, the shell effects seen in the
experiment are still without satisfactory explanation. A remaining possibility
is that the contact is governed by competing effects of the atomic packing and
the electronic free energy simultaneously. Similar mixed structures were also
found for Mg clusters~\cite{lyalin2003}.

In the total of our experiments, more than 90\% of the measured histograms for
all six samples at room temperature show a periodicity in $k_F R$, but the
observed frequencies in the total conductance interval vary in a range of 1 to
3.5~$(k_F R)^{-1}$. In 50\% of the histograms we obtained a value around
1.7~$(k_F R)^{-1}$ for lower conductance values. At higher conductance values
there is almost always a crossover to higher frequencies. The frequency at this
second conductance interval varies strongly and often there are multiple
frequencies superimposed. This indeed makes it likely that the shell effect is
the result of multiple properties of the metal.

Another feature of the histogram in \fref{fig:hit}, that is even more
remarkable, is the absence of plateaus below 5~G$_0$. This minimal value is
much higher than for all other metals studies with this
technique~\cite{mares2005,mares2007,mares2004}. Below this threshold value
we only found a smooth exponential decrease in conductance, typical for
tunneling behaviour. In principle, this suggests that smaller contacts are not
stable. However, from the analysis of the data, we cannot exclude that a
metal-to-insulator transition is taking place. As we explained in the
introduction, as the contact radius is reduced, the overlap between
the \textit{s} and \textit{p} states of the atoms in the constriction
decreases. Such a decrease could potentially lead to the opening of a
gap in the Mg density of states. In the next section we present a
theoretical analysis to elucidate this issue.

\section{Theory}\label{sec:theory}

\begin{figure}
\centerline{\includegraphics[width=4.5in]{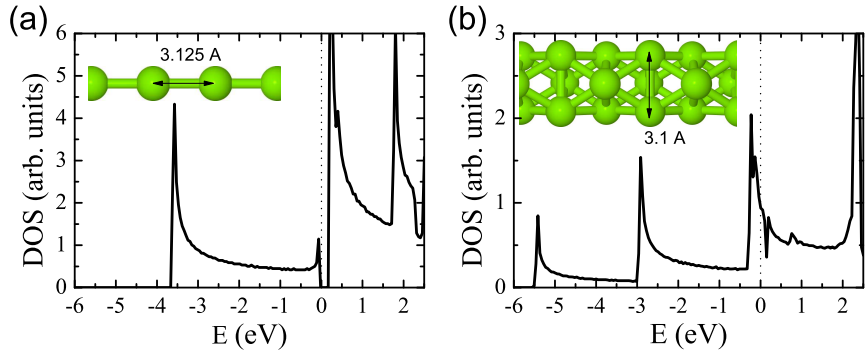}}
\caption{Density of states (DOS) of Mg wires calculated with
CASTEP~\cite{CASTEP}
for: (a) a linear chain with an optimal interatomic distance of 3.125~\AA; and
(b) an infinite wire with a 3 atoms section. The wire is built up along the
$c$-axis keeping the bulk symmetry as shown in the inset. The interatomic
distance that minimizes the total energy is 3.1~\AA.}
\label{PW_DOS}
\end{figure}

The numerous existent theoretical results on Mg clusters (see
reference~\cite{lyalin2003} and references therein) cannot be used
to resolve the problem of whether or not a Mg point contact can become
insulating. A metal-to-insulator transition is only well-defined for
truly infinite systems. Thus, in order to shed light on this problem,
we have studied the stability and electronic structure of a series of Mg
infinite wires with small coordination numbers. To be precise, we have
carried out density functional theory (DFT) calculations of the structural
and electronic properties of Mg wires ranging from the smallest possible
coordination, two in the case of a chain, up to 6. In the insets of
\fref{PW_DOS} we show two examples of the configurations of the
studied wires. In the upper panel one can see a linear chain, while the
lower one shows an infinite wire with a cross section of 3 atoms
(coordination number equals 6) grown along the $c$-axis of a hcp bulk
structure and keeping the bulk relative positions.

For these ab initio calculations we have used a standard implementation of
DFT~\cite{hohenberg1964,kohn1965} with a plane waves basis set and ultrasoft
pseudopotentials~\cite{vanderbilt1990}. The Perdew-Burke-Ernzerhof approach
(PBE)~\cite{perdew1996} has been chosen for the exchange-correlation
contribution. The calculations were performed with the CASTEP v4.2
code~\cite{CASTEP}. The plane wave cutoff used (375 eV) in all our
calculations assures well-converged structural and electronic properties.
Convergence criteria for the atomic relaxations involved in the different
calculations are: 0.01 eV/\AA~ for the mean value of the forces, 0.001 \AA~
for the atomic positions and 10$^{-6}$ eV for the total energy. We have
used a cubic supercell of 4 atoms for the linear chain and 2 atomic layers
for the wire adding a vacuum of 10 \AA~in the directions perpendicular to
the wires. We have optimized, minimizing the system energy, the interatomic
distances of the wires keeping the original symmetry with a Monkhorst-Pack (MP)
$\vec{k}$-sampling mesh of $1 \times 1 \times 64$ \cite{monkhorst1976}. The DOS
has been calculated with a mesh of $1 \times 1 \times 1024$.

Turning to the results, in the two systems shown in \fref{PW_DOS}, we have found
interatomic distances (3.1 \AA) slightly smaller than the bulk value (3.2 \AA)
in accordance with calculations of Mg clusters~\cite{lyalin2003}. In
\fref{PW_DOS}(a) we show the DOS of the linear chain for the optimized
interatomic distance~\footnote{We have studied the stability of the linear
chain allowing the 4 atoms of the supercell to relax starting form a zig-zag
configuration. For cell lengths larger than 10.5 \AA \ the final configuration
is a lineal chain slightly distorted, with a energy minimum corresponding to an
interatomic distance of 3.125 \AA \ and a zig-zag angle of 177$^\circ$. The
stable solution for smaller supercell lengths is a compressed zig-zag chain
where the atoms have 4 nearest-neighbors with an interatomic distance of 3.1
\AA.}. A small gap of 0.3 eV is observed; consequently the linear chain does not
show metallic behaviour. However, even for this wire with the lowest
coordination number, the broadening due to the 3\textit{s}-3\textit{p}
hybridization has nearly closed the gap. Indeed, we have found that all the
wires with larger coordination than the linear chain are metallic. The DOS of a
wire with a cross section of 3 atoms is shown in \fref{PW_DOS}(b) and does not
exhibit any gap. In fact, up to 5 different bands cross the Fermi level. We have
investigated whether the application of some additional stress could modify
these results, but we have found that even increasing the interatomic distances
by a 10\%, the wires remain metallic. So, in short, from this analysis we do not
expect the formation of insulating Mg wires in the last stages of the breaking
of Mg contacts.

The results above show that low-coordinated Mg structures are in principle
stable, which suggests that the formation of few-atom contacts might be
possible, at least at low temperatures. Moreover, since these structures are
metallic, the following natural question arises: What is the expectation
for the linear conductance of the smallest Mg contacts? In order to answer this
question we have computed the conductance of Mg one-atom contacts within the
Landauer formalism. For this purpose, we have combined the tight-binding
parameterization of reference~\cite{gotsis2002} with non-equilibrium Green's
functions techniques. This approach has been very successful describing the
transport properties of a great variety of metallic atomic-sized
contacts~\cite{brandbyge1999,dreher2005,viljas2005,pauly2006}. We proceed to
briefly explain it. In this approach the electronic structure of the atomic
contacts is described in terms of the following Hamiltonian written in a
nonorthogonal local basis
\begin{equation}
\label{Ham}
\hat{H} = \sum _{i\alpha ,j\beta ,\sigma } H_{i\alpha ,j\beta}
\hat c_{i\alpha ,\sigma }^{\dagger} \hat c_{j\beta ,\sigma } ,
\label{eq:nonorthogonalH}
\end{equation}
where $i$ and $j$ run over the atomic sites, $\alpha $ and $\beta $ denote
different atomic orbitals, $\sigma$ is the spin and $H_{i\alpha ,j\beta}$ are
the on-site energies ($i=j$) or hopping elements ($i\neq j$). Additionally, we
need the overlap integrals $S_{i\alpha ,j\beta }$ of orbitals at different
atomic
positions. We take these matrix elements from the tight-binding parameterization
of
reference~\cite{gotsis2002}, which is designed to accurately reproduce the band
structure of bulk materials. The atomic basis is formed by 9 valence orbitals,
namely the \textit{s}, \textit{p}, and \textit{d} orbitals which give rise to
the
main bands around the Fermi energy. In this parameterization both the hoppings
and the overlaps are functions of the atomic distances, which have a cutoff
radius that encloses up to 13 nearest-neighbour shells. Finally, in order to
take into account the low coordination in the smallest atomic contacts, we
impose local charge neutrality through the self-consistent variation of the
on-site energies of the atoms in the constrictions.

With the help of Green's functions techniques, one can translate the
information on the electronic structure contained in the Hamiltonian of
equation~\eref{Ham} into the conductance of these atomic junctions (see
references~\cite{dreher2005,pauly2006} for details). As explained in the
introduction, this low-temperature conductance adopts the form of the Landauer
formula
\begin{equation}
G = \mbox{G}_{0} T\left(E_{F}\right) = \mbox{G}_{0} \sum _{n}\tau_{n} ,
\label{eq:conductanceG}
\end{equation}
where G$_{0}=2e^2/h$ denotes the quantum of conductance, $E_{F}$ the Fermi
energy, and $\tau_{n}$ the transmission of the $n$-th transmission
eigenchannel at $E_{F}$.

\begin{figure}
\centerline{\includegraphics[width=5in]{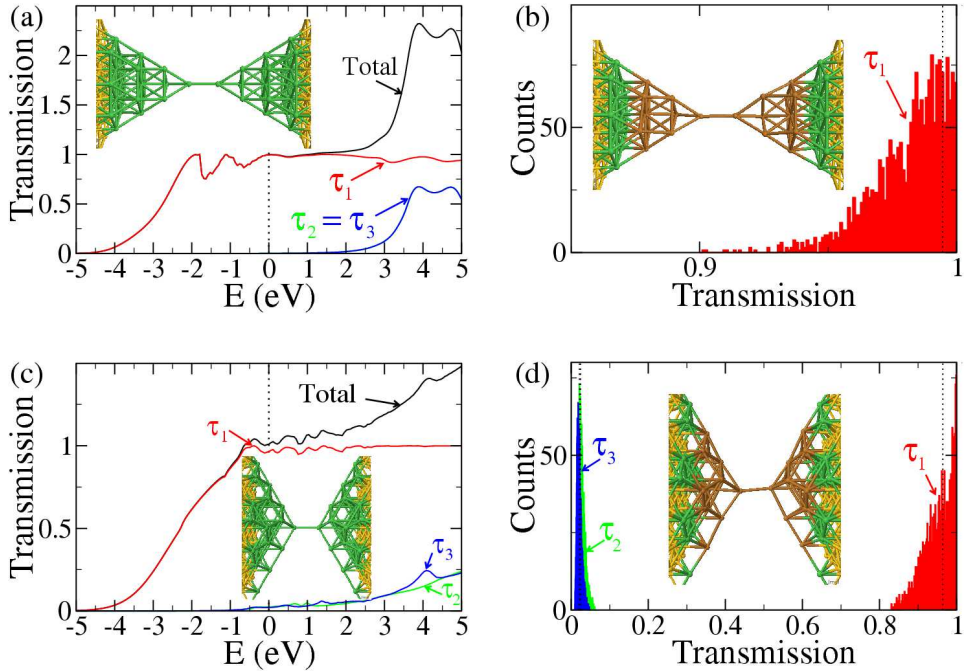}}
\caption{(a) Total transmission and transmission coefficients as a function
of energy for the contact shown in the inset. This geometry is grown along
the [0001] direction ($c$ axis), it contains a dimer in its central part and
the interatomic distances are fixed to the bulk values. The yellow atoms
correspond to atoms in the semi-infinite surfaces that are used to model the
leads. The transmission coefficients at the Fermi energy (set to zero and
indicated with a dotted vertical line) are $\tau_1 = 0.993$, $\tau_2 =
\tau_3 = 0.76 \times 10^{-3}$ and the total conductance is $\sim 0.995$~G$_0$.
(b) Histograms of the transmission coefficients for 2000 disorder realizations
(see text) of the one-atom contact of the inset of panel (a). The brown atoms
are those which have been randomly displaced. The vertical dotted lines
indicate the values of the transmission coefficients for the ideal geometry.
The transmission values of the second and third channels remain below $2
\times 10^{-3}$ and are not shown. (c) The same as in panel (a), but for the
contact grown along the [11$\bar{2}$0] direction ($a$-axis) shown in the inset.
The total conductance is $\sim 1.01$~G$_0$ with $\tau_1 = 0.9640$, $\tau_2 =
0.024$, $\tau_3 = 0.021$ and $\tau_4 = 0.3 \times 10^{-3}$. (d) The same as in
panel (b) for the contact in the inset of panel (c).}
\label{cond-thy}
\end{figure}

We now turn to the analysis of the results for the linear conductance of
some ideal, and yet plausible, one-atom geometries, which have been chosen
to simulate what may happen at the last conductance plateau before the
rupture of the nanowires. In the inset of \fref{cond-thy}(a) we show
an example of an one-atom contact grown along the [0001] direction
($c$-axis) which contains a dimer in its central part. Different molecular
dynamics simulations of atomic contacts of various metals have suggested
that this type of geometry is realized very frequently at the last
plateau~\cite{dreher2005,pauly2006,jelinek2003}. This particular geometry
is constructed starting with the dimer and choosing the nearest-neighbours
in the next layers. Finally, the leads are modelled as infinite surfaces
grown along the same direction (yellow atoms in the inset of
\fref{cond-thy}(a)). In this ideal case, all interatomic distances are fixed to
their bulk values, which is justified by the ab initio calculation of Mg
clusters~\cite{lyalin2003} and our results for infinite wires presented above.
The total transmission and the individual transmission coefficients, $\tau_i$,
for this contact are shown in \fref{cond-thy}(a) as a function of energy. The
first thing to notice is the fact that the system is metallic, in accordance
with our expectations based on the DFT results described above. Furthermore, as
one can see, the conductance, which is determined by the transmission at $E_F$,
is very close to 1~G$_0$ and it is completely dominated by a single fully open
channel. The second and third channels, which are degenerate due to the symmetry
of the contact, have transmissions below $10^{-3}$. One can get a deeper insight
into these results and, in particular, into the nature of the conduction
channels by analyzing the local density of states projected onto the different
orbitals of the two central atoms (not shown here). Such analysis indicates that
the dominant channel is formed by a symmetric (bonding) combination of the
\textit{s} and \textit{p}$_z$ orbitals of the central atoms ($z$ is the
transport direction), while the \textit{p}$_x$ and \textit{p}$_y$ orbitals are
responsible for the second and third channels.  A fourth channel, which in this
case has a transmission below $10^{-5}$, is formed by the antisymmetric
combination of the \textit{s} and \textit{p}$_z$ orbitals. Such anti-bonding
combination is basically orthogonal to incoming states (from the leads) and
therefore does not contribute significantly to the transport. So in short, these
results resemble very much what happens in the case of Zn one-atom
contacts~\cite{hafner2004}, and also in the case of the final stages of the last
plateau of Al contacts~\cite{scheer1997,cuevas1998}. This is, after all, quite
reasonable since in all these cases the electronic structure at the Fermi energy
is governed by the \textit{s} and \textit{p} orbitals.

Since we do not exactly know the growth direction of the Mg atomic contacts,
we have studied the conductance of geometries with different crystallographic
orientations. In \fref{cond-thy}(c) we show another example of a dimer
contact, but this time grown along the [11$\bar{2}$0] direction ($a$-axis).
Notice that the total conductance and the transmission coefficients are
similar to those of the [0001] case, the main difference being the larger
values for the second and third channels and the lifting of their degeneracy.
While the lack of degeneracy reflects the lower symmetry of the contacts in
[11$\bar{2}$0] direction, the larger transmission values can be attributed
to the larger apex angle of those contact geometries and in consequence a
stronger coupling of the dimer atoms and next layers.

In principle, the contact geometries should be determined from molecular
dynamic simulations, but this is computationally very demanding. Instead,
and in order to test the robustness of our results, we have studied the
role of disorder in the atomic positions. For this purpose, starting from
the ideal geometries of \fref{cond-thy}(a) and (c), we have changed
randomly the positions of the atoms in the constriction region (those
highlighted in brown in \fref{cond-thy}(b) and (d)) with a maximum
amplitude of $\pm$5\% of the nearest-neighbour distance. Then, we have
computed the total transmission and the transmission coefficients of the
disorder geometries and the results are shown in the form of histograms
in \fref{cond-thy}(b) and (d). As one can see, in both cases the
conductance is still dominated by a single channel that is almost fully
open. Therefore, these results confirm our basic conclusion, namely the
fact that a Mg one-atom contact is expected to have a conductance close
to 1~G$_0$ dominated by a single channel.

\section{Low temperature results}\label{sec:tlow}

\begin{figure}
\centerline{\includegraphics[width=5in]{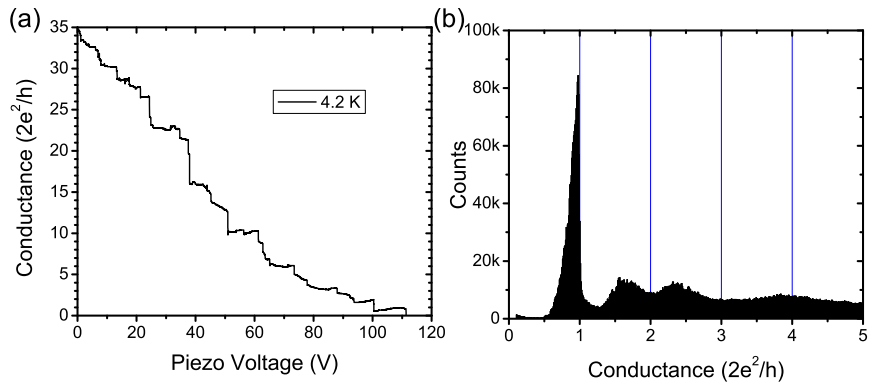}}
\caption{Figure comparable to \fref{fig:hit}, but for a measurement
performed at liquid helium temperature. Panel (a) shows the conductance
evolution measured at a bias voltage of 100~mV. Panel (b) shows a conductance
histogram resulting from $3 \times 10^3$ of these traces. The resolution of this
histogram was set to 115 bins per G$_0$.}
\label{fig:lowt}
\end{figure}

In order to test this latter conclusion of our theoretical analysis, we
repeated the experiments at low temperatures. Here we broke the contact
by ramping the piezo element continuously with a speed of $1.8\times
10^3$~V/s (corresponding to $1.0\times 10^2$~nm/s electrode separation).
This resulted in traces as the one shown in \fref{fig:lowt}(a). The plateaus
in conductance appear more structured and the jumps during the mechanical
reconfigurations are smaller than those measured at room temperature. In the
conductance histogram, presented in \fref{fig:lowt}(b), this results in a flat
distribution at higher conductance values. The shell structures found at room
temperature are thus absent at low temperatures, as we expected.

When focusing on the trace at low conductance values, the staircase of
plateaus indeed continues down to values below 5~G$_0$. The histogram,
the shape of which was reproduced over a set of ten different samples,
shows its lowest conductance peak close to 1~G$_0$. For a minority of
the histograms a shoulder down to values as low as 0.8~G$_0$ is seen.
In rare cases this shoulder even grows to form the primary peak.
Although the conductance does not reveal information on the individual
values of transmission, $\tau_i$, of the electron states, the strong
asymmetry and the closeness of the peak to 1~G$_0$ do suggest that its
conductance is given by only one, almost fully opened, channel. For
multiple channels the value of 1~G$_0$ would not form a fundamental limit.

This result agrees nicely with our calculations for the conductance of Mg
single-atom contacts. The properties of Mg at low temperatures are therefore
very close to the results obtained previously for Zn~\cite{hafner2004}. The
most noticeable differences are the absence of data points between 0.5~G$_0$
and tunneling and the higher relative intensity of the peak close to 1~G$_0$
in the case of Mg. The presence of this peak indicates that the instability
of small point contacts at room temperature is not related to a
metal-to-insulator
transition.

\section{Conclusions}\label{sec:conc}

In conclusion, magnesium contacts at room temperature demonstrate shell
effects at multiple frequencies. The most frequent and intense frequency
of 1.7 $(k_FR)^{-1}$ did not match with either the expected frequency for
atomic packing or the previously calculated frequencies for electronic
shell structures. This value therefore remains without satisfactory
explanation. A possibility is that both effects play an important role
at the same time.

The histograms at low temperatures exhibit a first peak close to 1~G$_0$.
The strongly asymmetric shape of the peak, with a small weight above 1~G$_0$
suggests this conductance is dominated by a single channel. Our calculations
confirm this and identify this channel to be a symmetric combination of the
\textit{s} and \textit{p}$_z$ orbital of the central Mg atom. This behaviour is
similar to the results obtained for Zn~\cite{hafner2004}, although the tendency
for the channel to be fully open is stronger.

At room temperature contacts of the diameter of a single atom are absent.
The instability of these smaller contacts is not caused by a metal-to-insulator
transition at lower coordination. From the appearance of the peak at 1~G$_0$
at low temperatures, we can conclude that the metal-insulator transition is
absent even in the smallest of contacts. Future experiments at intermediate
temperatures may prove valuable in understanding these different behaviours. Our
first principle calculations suggest that an infinite linear chain of Mg could
be an insulator, but they also show that as soon as the coordination is larger
than 2, as in any realistic contact at low temperature, the system is metallic.

\ack

We would like to thank Gijs van Dorp and Xin-Zhou Liu for their assistance with
the measurements. This work is part of the research programme of the `Stichting
voor Fundamenteel Onderzoek der Materie (FOM)' (RHMS, AIM and JMvR), which is
financially supported by the `Nederlandse Organisatie voor Wetenschappelijk
Onderzoek (NWO)'. MH acknowledges financial support from the KHYS and the DFG
within the CFN and PP and JCC are supported by the Spanish MICINN under the
Grants \mbox{MAT2008-02929/NAN} and \mbox{FIS2009-04209}, respectively.

\end{document}